\documentclass{aa} 
\usepackage{hyperref}
\usepackage{natbib}
\bibpunct{(}{)}{;}{a}{}{,} 

\usepackage{graphicx}
\usepackage{txfonts}
\begin{document}

   \title{The blue straggler population of the old open cluster Berkeley 17}
   \titlerunning{Blue stragglers in Berkeley 17}

   \author{Souradeep Bhattacharya\inst{1} \and
          Kaushar Vaidya\inst{2} \and
          W. P. Chen\inst{3} \and
          Giacomo Beccari\inst{1}
          }
   \institute{European Southern Observatory, Karl-Schwarzschild-Str. 2, 85748 Garching, Germany \\ 
   \email{sbhattac@eso.org} \and
   Department of Physics, Birla Institute of Technology and Science, Pilani 333031, Rajasthan, India \and
   Graduate Institute of Astronomy, National Central University, 300 Jhongda Road, Jhongli 32001, Taiwan
             }

   \date{Received: October 17, 2018; Accepted: February 15, 2019}

 
\abstract
{Blue Straggler Stars (BSSs) are observed in Galactic globular clusters and old open clusters.  The radial distribution of BSSs has been used to diagnose the dynamical evolution of globular clusters.  For the first time, with a reliable sample of BSSs identified with Gaia DR2, we conduct such an analysis for an open cluster. }
{We identify members, including BSSs, of the oldest known Galactic open cluster Berkeley 17 with the Gaia DR2 proper motions and parallaxes. We study the radial distribution of the BSS population to understand the dynamical evolution of the cluster.}
{We select cluster members to populate the colour magnitude diagram in the Gaia filters. Cluster parameters are derived using the brightest members. The BSSs and giant branch stars are identified, and their radial distributions are compared.  The segregation of BSSs is also evaluated with respect to the giant branch stars using the Minimum Spanning Tree analysis.}
{We determine Berkeley 17 to be at $3138.6^{+285.5}_{-352.9}$ pc. We find 23 BSS cluster members, only two of which were previously identified.  We find a bimodal radial distribution of BSSs supported by findings from the MST method.}
{The bimodal radial distribution of BSSs in Berkeley 17 indicates that they have just started to sink towards the cluster center, placing Berkeley 17 with globular clusters of intermediate dynamical age. This is the first such determination for an open cluster.}


\maketitle
%

\section{Introduction}
In the colour magnitude diagrams (CMDs) of globular clusters and old open clusters, blue straggler stars (BSSs) are observed to be brighter and bluer than the main sequence turn-off lying along an extrapolation of the main sequence \citep{sandage53}.  {Mass transfer in a binary system \citep{McCrea64} and direct stellar collisions \citep{Hills76} have been considered to be the possible physical mechanism for BSS formation.}  {BSS populations and their formation scenarios in globular clusters have been the subject of many studies to understand their nature and evolution in the context of known stellar evolution processes \citep[eg.][]{Davies04,Ferraro09,Sills13}}. 

Irrespective of their formation mechanism, BSSs are more massive than the giant branch stars and their radial distribution has been used as ``dynamical clocks'' for globular clusters to measure the level of dynamical evolution of the system. \citet{Ferraro12} showed that based on the BSS radial distribution with respect to that of the light (or reference stars), globular clusters can be classified into three families: \textit{Family I} globular clusters have a flat distribution and are ``dynamically young''; \textit{Family II} globular clusters have a bimodal distribution, with a peak at the cluster center, a minimum at intermediate radii, a rise in the external regions \citep[eg.][]{Dal09,Bec12}, and are evolving under efficient dynamical friction; \textit{Family III} globular clusters have a central peak followed by a monotonically decreasing trend and are ``dynamically old''.

In the case of old open clusters, while a large number of BSSs have been reported \citep{al07}, their numbers remain sparse with uncertain membership except for the very close open clusters (eg. NGC 188; \citeauthor{mg09} \citeyear{mg09}, M67; \citeauthor{Bertelli18} \citeyear{Bertelli18}). It has thus not been possible to study BSSs in open clusters as thoroughly as in globular clusters, since their radial distribution remains elusive with uncertain membership. Possible formation mechanisms have been examined only for the very close open clusters. However, the sparse nature of open clusters make them ideal laboratories in which to study the nature and formation of BSSs with spectroscopy.

Berkeley 17 (RA=05:20:37, DEC=+30:35:12, J2000, hereafter Be17) is located near the Galactic anti-center at a distance of $\sim2.7$ kpc \citep{Phelps97}.  {With a metallicity ${\rm[Fe/H]}\approx-0.33$ \citep{Friel02} and an age of $\sim10$ Gyr \citep{kal94,Phelps97,sal04,kru06}, it is the oldest known open cluster. While \citet{Bragaglia06} found a slightly lower age of 8.5--9~Gyr, these authors did not rule out an older age up to $\sim12$~Gyr.} It is mass-segregated and has a distinct tidal tail structure \citep{Chen04,Bh17} possibly resulting from influence by the Perseus arm of the Galaxy. It was also known to be rich in BSSs with 31 candidates by \citet{al07} though \citet{Bh17} showed that around half of those might be field stars.  {Being close to the galactic plane, Be17 has a significant field contamination which in the CMD occupy the same region as the BSSs. Thus, membership determination is essential to study the BSSs in Be17.}

With the second Data Release of Gaia \citep[hereafter GDR2]{Gaia18}, accurate proper motions and parallaxes have become available to allow us to identify the bright members of Be17, including the BSSs. We can hence for the first time use the BSS population to probe the dynamical state of an old open cluster, following the method suggested by \citet{Ferraro12}. It is important to identify if BSSs are efficient test-particles to infer the dynamical state of a stellar system like an open cluster, which offers a completely different environment (in terms of total mass, formation history and stellar density) as compared to globular clusters. We describe the data used in Sect~\ref{data} to determine the cluster members in Sect~\ref{memb}. We derive the properties of the cluster in Sect~\ref{prop} and characterize the BSS population and its radial distribution in Sect~\ref{bss}. Finally, we end with a discussion in Sect~\ref{diss}.

\section{Data description}
\label{data}
GDR2 provides position, trigonometric parallax, and proper motion as well as photometry in three broad-band filters (G, GBP, and GRP) for more than a billion stars. It also provides spectroscopic radial velocities for the brightest stars. The astrometric solution is described in \citet{Lin18}, the photometric content and validation is described in \citet{Evans18}, while the spectroscopy is described in \citet{Katz18}.  {From the known center of Be17, its core region and tidal tails are within a radius of $\sim11\arcmin$ \citep{Bh17}. So we utilized GDR2 data of 8357 stars within a slightly larger 15$\arcmin$ radius from the known center of Be17 for our analysis.}  

\begin{figure}[t]
	\centering
	\includegraphics[width=0.9\columnwidth,angle=0]{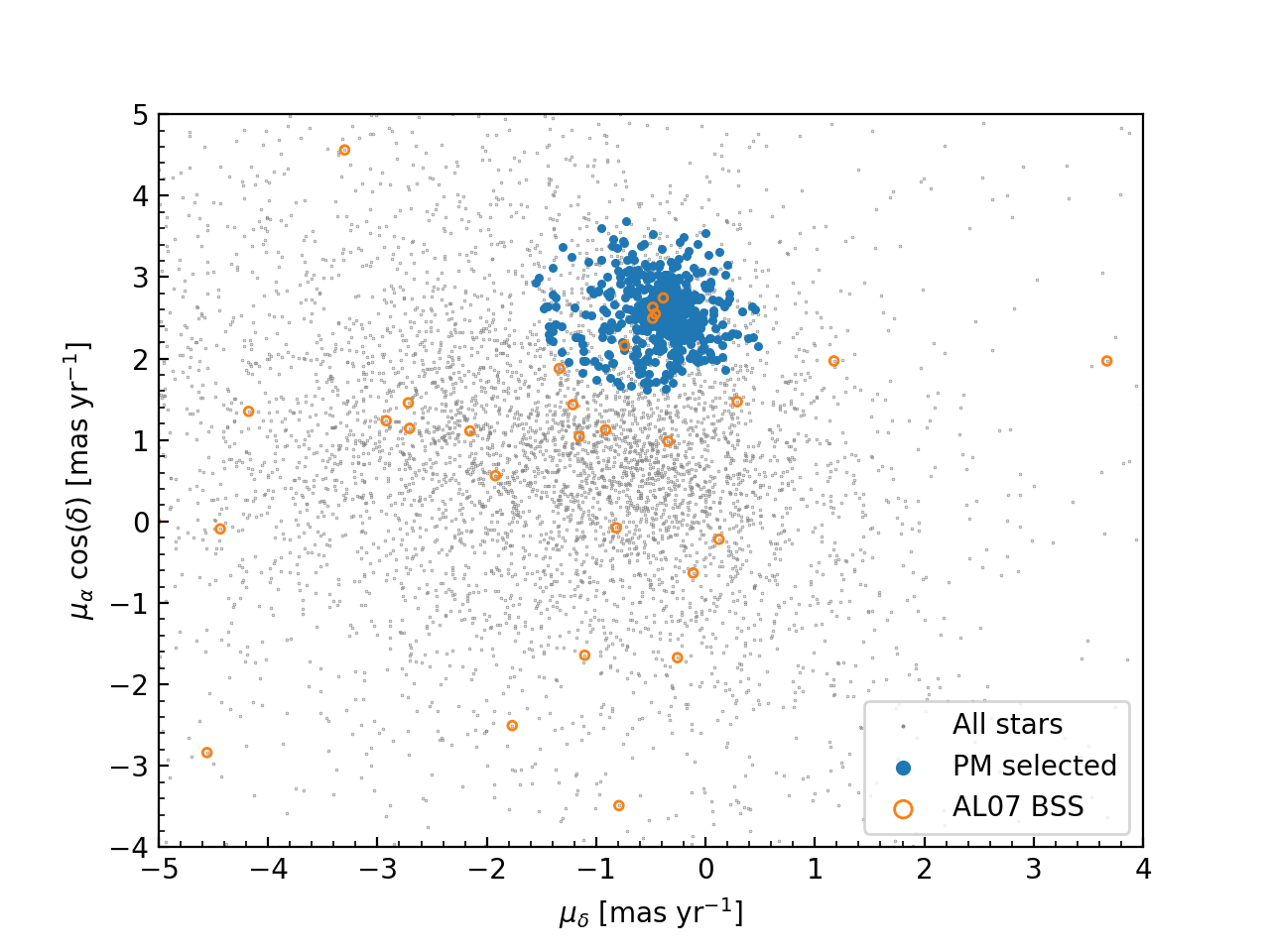}
	\includegraphics[width=0.9\columnwidth,angle=0]{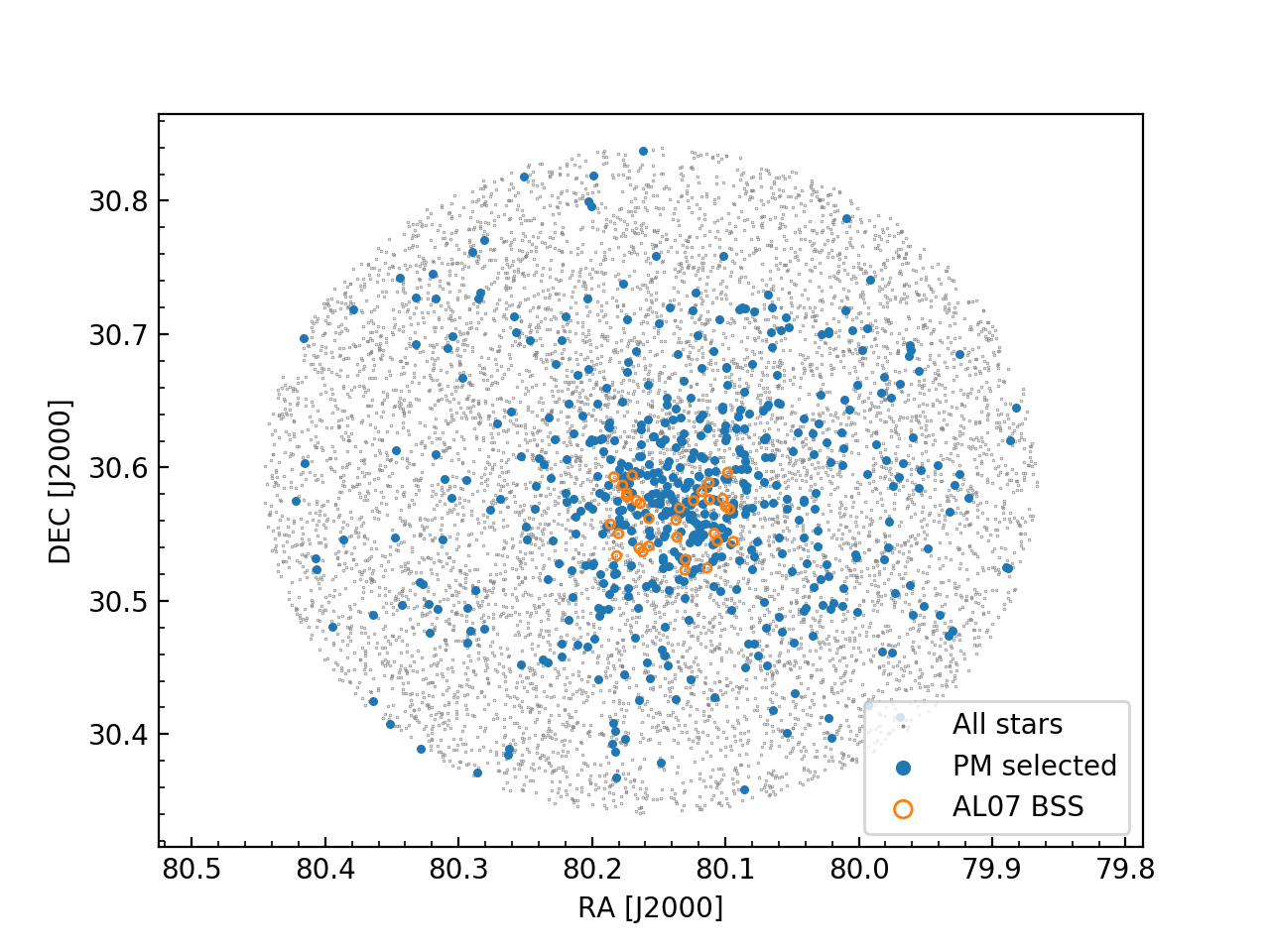}
	\includegraphics[width=0.9\columnwidth,angle=0]{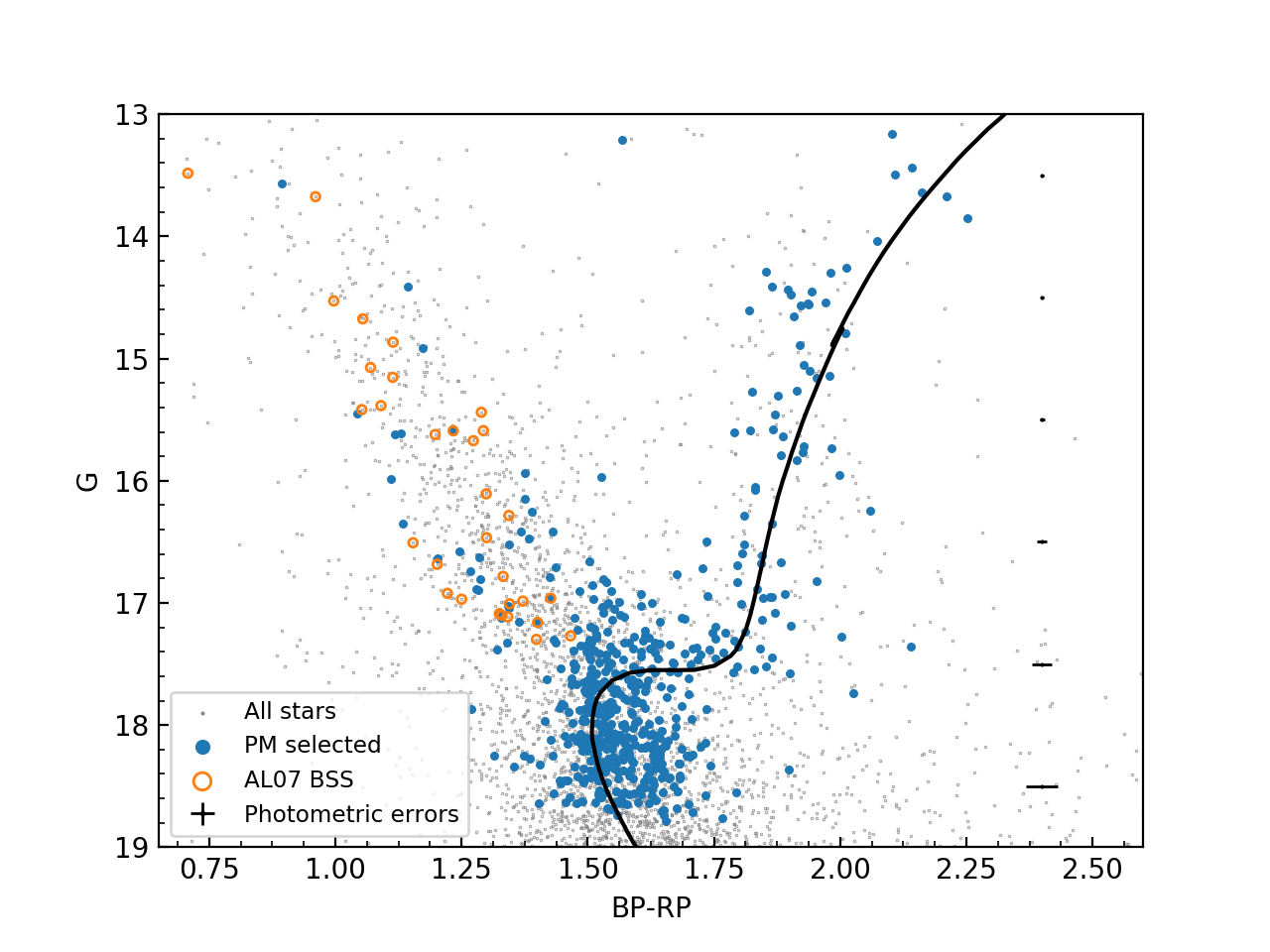}
	\caption{ {All the sources with 15$\arcmin$ from the known center of Be17 are shown in grey, the proper motion selected members are shown in blue, and the BSS candidates from \citet{al07} are shown in orange. [Top] The proper motion of the stars. [Middle] The spatial distribution of the stars. [Bottom] The CMD for the stars where the isochrone, shown with a black solid line, corresponds to the adopted age and metallicity of Be17. The photometric errors are shown on the right in black. The errors in magnitude are negligible.}}
	\label{Fig:pmsel}
\end{figure}

\section{Membership determination}
\label{memb}
 {We utilize a procedure similar to \citet{Yen18} to identify Be17 cluster members based on the GDR2 proper motions and parallaxes. The GDR2 counterparts of previously known members are first identified. To identify possible members of Be17 with GDR2, we select stars within 3 times the standard deviation of the mean proper motion of the known members. We only select those sources whose proper motion errors are within 0.5~mas~yr$^{-1}$, where mas stands for milli-arcsecond, to ensure accuracy. Of the sources thus selected, we further consider only those as members whose trigonometric parallax values are within 1.5 times the standard deviation of the known members.}

 {\citet{Scott95} had identified 13 giant stars as members of Be17 based on their location on the CMD and radial velocities. Of these, only 12 were confirmed as members by \citet{Friel02} based on their metallicity, [Fe/H], which was determined for each star from spectroscopic indices that primarily measured \ion{Fe}{I} and \ion{Fe}-peak blends.} We identify the GDR2 counterparts to these stars as the nearest neighbours within 0.2$\arcsec$. We find the mean proper motion of these 12 giants as $\mu_{\rm \alpha} \cos\delta = 2.71 \pm 0.07$~mas~yr$^{-1}$, $\mu_{\rm \delta} =-0.48 \pm 0.13$~mas~yr$^{-1}$.   {Within errors, its close to the previously adopted uncertain value of $\mu_{\rm \alpha} \cos\delta = 3.60 \pm 3.34$~mas~yr$^{-1}$, $\mu_{\rm \delta} =-3.62 \pm 2.27$~mas~yr$^{-1}$ obtained by  \citet{dia14} from proper-motion identified members from UCAC4 (the fourth U.S. Naval Observatory CCD Astrograph catalog).} The increased accuracy is a testament of the quality of the GDR2 data. We also find the mean trigonometric parallax of the 12 giants as $\overline{\omega} = 0.3 \pm 0.03$~mas. 

\begin{figure}[t]
	\centering
	\includegraphics[width=0.9\columnwidth,angle=0]{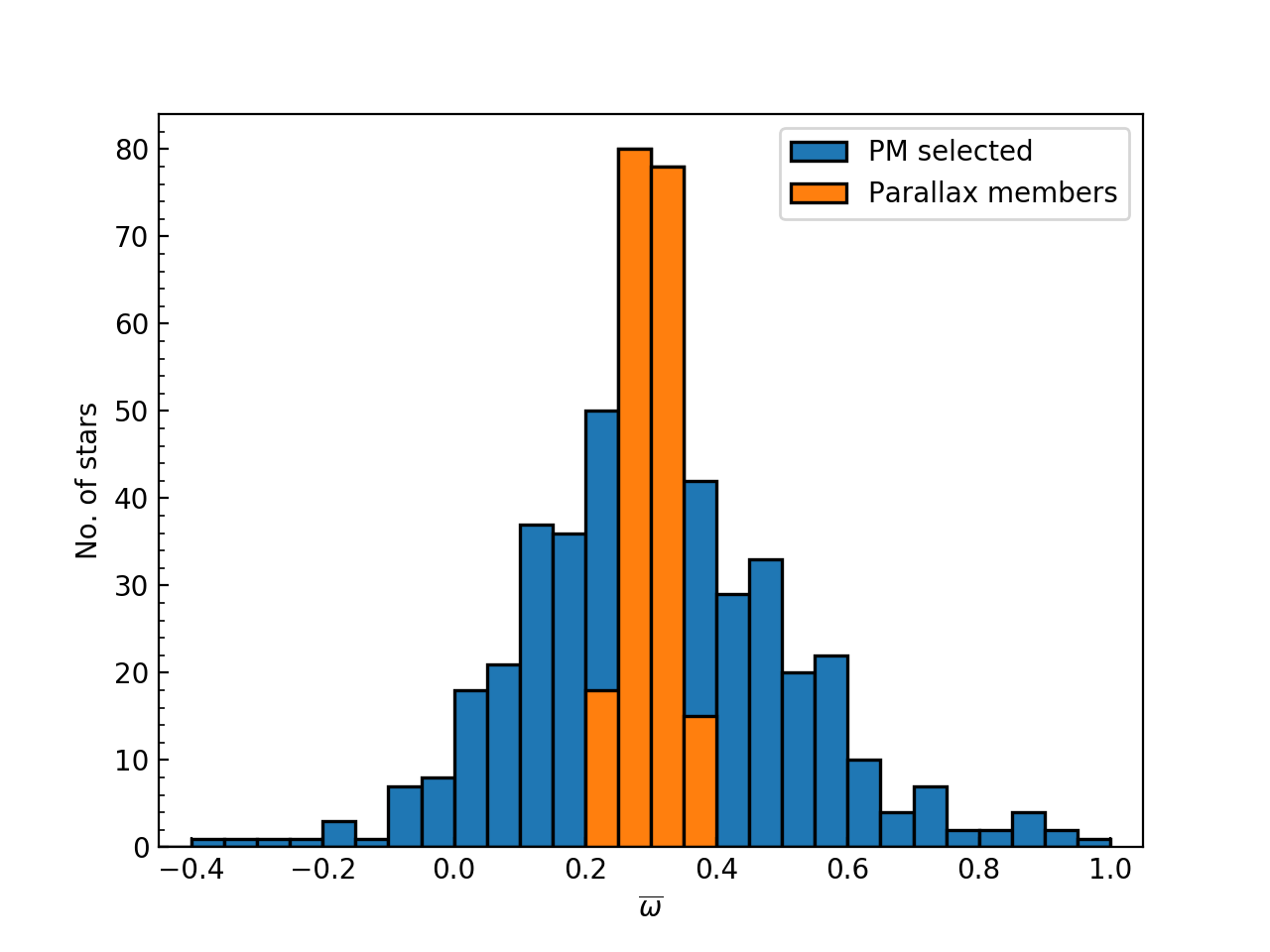}
	\includegraphics[width=0.9\columnwidth,angle=0]{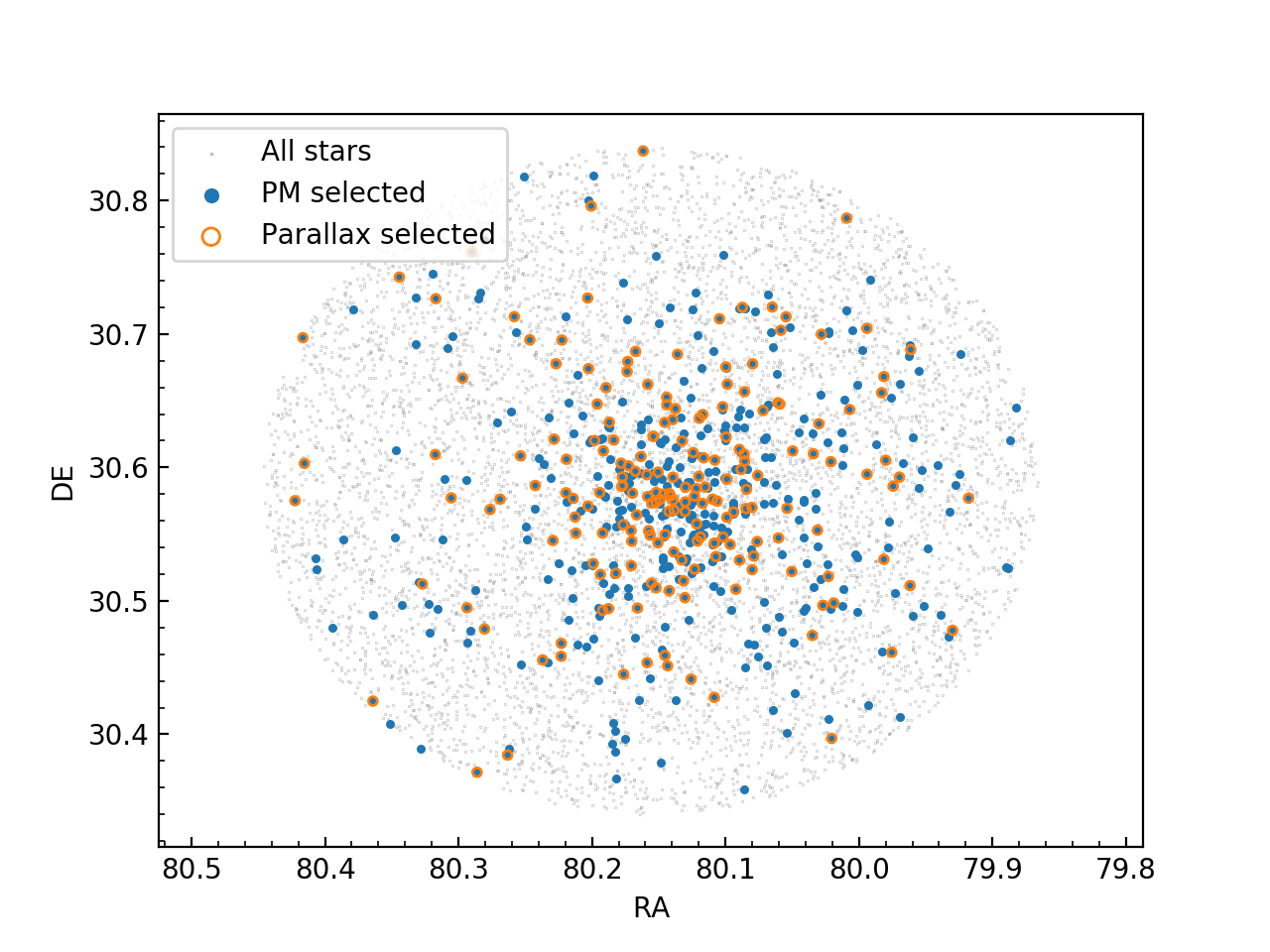}
	\includegraphics[width=0.9\columnwidth,angle=0]{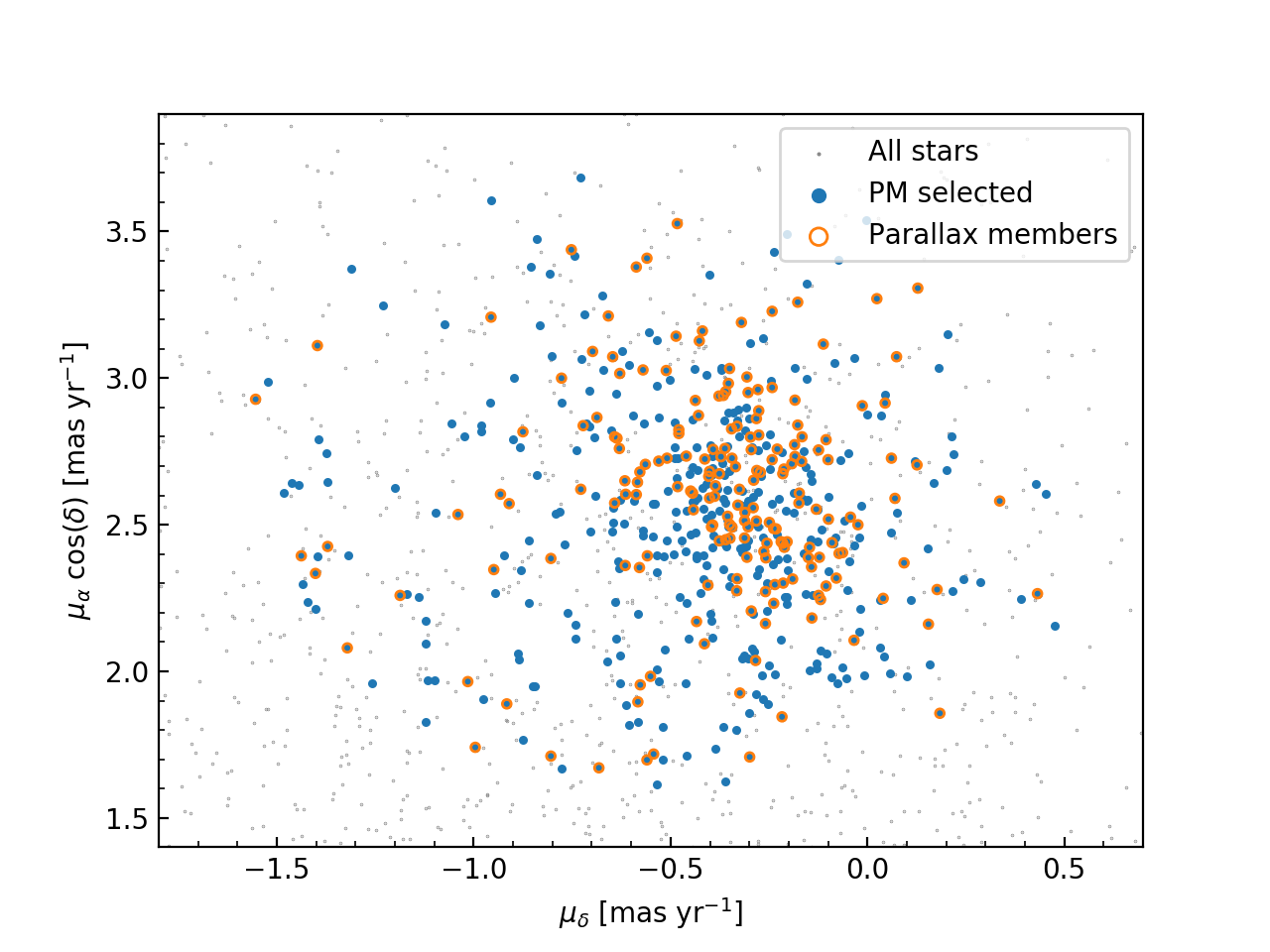}
	\caption{ {[Top] The proper motion selected members are shown with the blue histogram while the parallax selected members are shown with the orange histogram. [Middle] The spatial distribution of the stars with all the sources with 15$\arcmin$ from the known center of Be17 shown in grey, the proper motion selected members shown in blue, and the parallax selected members shown in orange. [Bottom] The proper motion of the same stars.}}
	\label{Fig:parasel}
\end{figure}

 {Proper motion selected members of Be17 are those within 3 times the standard deviation of the mean proper motion of the known giants whose errors on the $\mu_{\rm \alpha} \cos\delta$ and $\mu_{\rm \delta}$ are within 0.5~mas~yr$^{-1}$. Fig~\ref{Fig:pmsel} shows the 523 proper motion selected stars in blue while all the stars within 15$\arcmin$ from the center of Be17 are shown in grey. The BSS candidates identified by \citet{al07} are shown in orange. Only 5 of them are classified as members in the proper motion selection, as evident in the top panel of Fig~\ref{Fig:pmsel}. The middle-panel of Fig~\ref{Fig:pmsel} shows the spatial distribution of the stars where the proper motion selected members appear centrally concentrated with an apparent cluster halo. The stars at the outskirts of the spatial distribution may be non-members. From the CMD shown in the lower panel of Fig~\ref{Fig:pmsel}, it is evident that the proper motion selected members follow the isochrone (described later in Sect~\ref{prop}) suitable for Be17. The overlap of field stars in the region of the CMD occupied by BSSs is evident and this resulted in the misidentification of field stars as BSSs by \citet{al07}. To further remove field contamination, the membership determination may be polished by parallax selection of cluster members.}  

 {Of the proper motion selected members, only those are considered as cluster members whose trigonometric parallax values are within 1.5 times the standard deviation of the known giants.} With a more relaxed parallax selection, more faint sources with higher parallax uncertainties would have been selected. So in order to limit contamination from field stars, the stringent selection criteria in parallax was applied. We do not apply an error cut to the parallax selection to avoid being biased towards more nearby sources, which have less uncertainties in their parallax determination.  {The parallax selection is shown in the top panel of Fig~\ref{Fig:parasel}}. In total we identify 191 sources as members of Be17.  {The spatial distribution of stars shown in the middle panel of Fig~\ref{Fig:parasel} still shows that some of the identified cluster members are in the outskirts of the search area. The lower panel of Fig~\ref{Fig:parasel} shows that a majority of the parallax selected members showed a concentration in the proper-motion space slightly offset from the mean proper motion of the known giants, indicating that the mean proper motion of Be17 is slightly different from that of the known giants members. A few stars appear further away from this concentration but it is a significant improvement from just the proper motion selection. These may indeed be field contaminants and the membership selection may be further refined with the next Gaia data release. The membership determination is accurate enough to study the bright sources in Be17, including the BSSs.}


\section{Cluster properties}
\label{prop}
Of the identified members, six sources have GDR2 radial velocity information from low resolution spectrophotometry, giving a mean radial velocity, V$_{\rm rad} = -72.86 \pm 0.98$ km s$^{-1}$.  {This is close to V$_{\rm rad} = -84$ km s$^{-1}$ derived by \citet{Scott95} for their 12 giants within their estimated standard deviation, $\sigma_{\rm v}=11$ km s$^{-1}$. Three of these bright GDR2 sources are the counterparts of the known giants having IDs 4607, 0079 and 0099 in the catalog of \citet{Scott95}. Their individual radial velocities match within errors as \citet{Scott95} note that each of their individual radial velocity measurements may have uncertainty of $\sim10$ km s$^{-1}$.} For the six sources observed with spectrophotometry, GDR2 also provides extinction in the G-filter, A$_{\rm G}$, and reddening, $E(BP-RP)$, inferred using the Apsis-Priam system \citep{bj13}. We find a mean A$_{\rm G} = 1.514$ and mean reddening $E(BP-RP) = 0.7358$ for these six sources. 

 {Since reliable distances to the GDR2 sources cannot be obtained by simply inverting the parallax, \citet{bj18} use an inference procedure to obtain distances for each of the GDR2 sources accounting for the nonlinearity of the transformation and the asymmetry of the resulting probability distribution. The mean distance of Be17 is the mean of the individual source distances obtained by \citet{bj18} for the brightest cluster members having G < 15 mag.} It is obtained as $3138.6^{+285.5}_{-352.9}$ pc, in good agreement with that found by \citet{cg18} also using GDR2 but selecting cluster members with a different method. We use only the brightest members because the uncertainties on parallax are much larger for the fainter sources. The derived distance to Be17 is slightly farther than the previously known value of 2.7 kpc \citep{Phelps97}.

\begin{figure}[t]
	\centering
	\includegraphics[width=\columnwidth,angle=0]{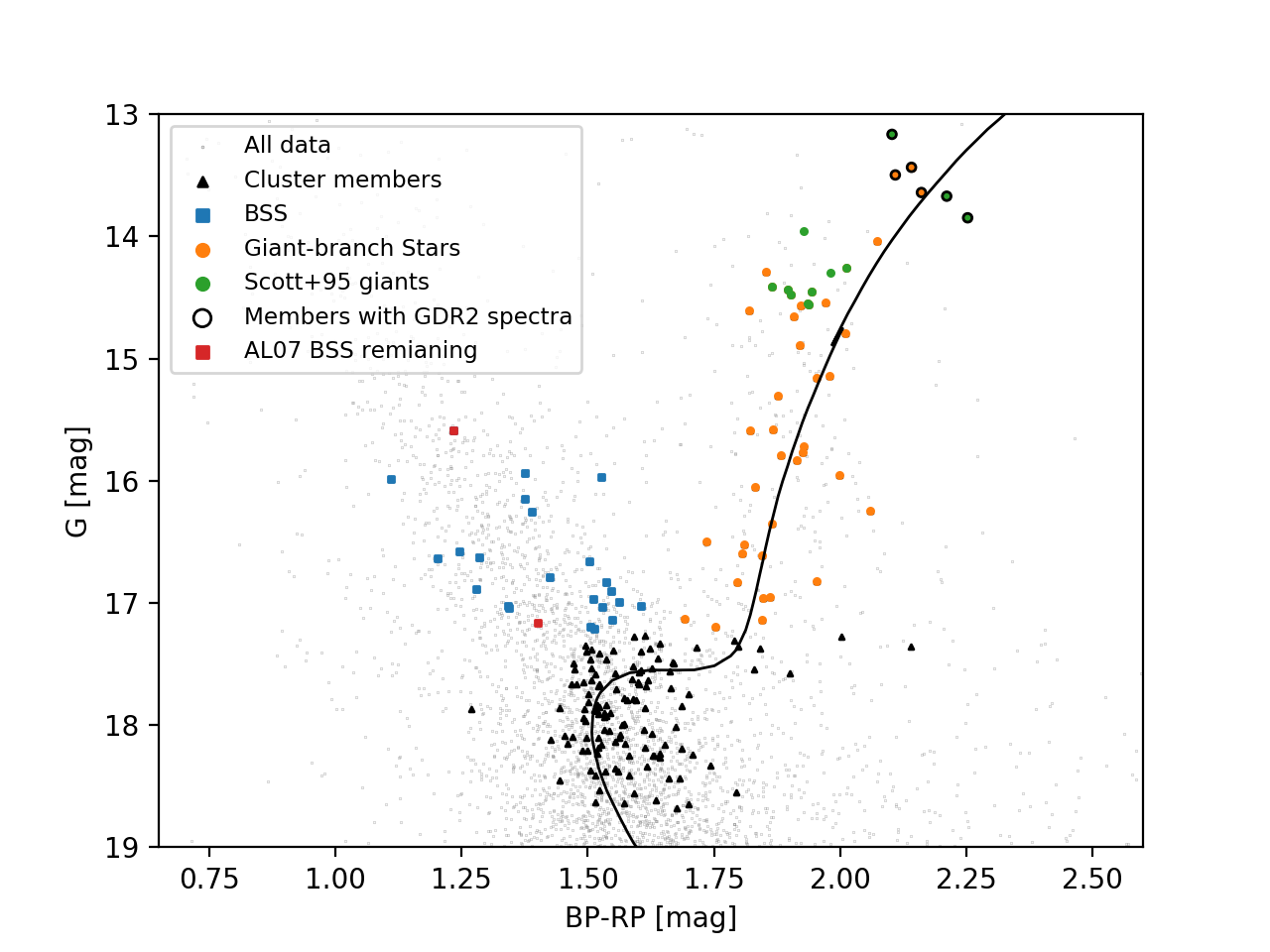}
	\caption{The CMD showing cluster members in black triangles while all sources in the search area are shown in grey.  {The cluster members along the giant branch are shown in orange while those shown as blue squares are the BSSs. The two dark red squares are BSSs identified by \citet{al07} and reidentified in this work. The 12 giants from \citet{Scott95} are shown in green and are re-identified as cluster members in this work. The 6 stars with spectrophotometric data from GDR2 are encircled in black.} The PARSEC isochrone, shown with a black solid line, corresponds to the adopted age and metallicity of Be17.}
	\label{Fig:cmd}
\end{figure}

For all the identified members, we obtain the CMD in the GDR2 colours shown in Fig~\ref{Fig:cmd}. The PARSEC stellar evolution isochrone\footnote{http://stev.oapd.inaf.it/cgi-bin/cmd} \citep{bre12,che14,marigo17} for the Gaia filters has been plotted corresponding to the 10 Gyr age of Be17, its metallicity of 0.007 ($[Fe/H] = -0.33$), the obtained distance, and the obtained A$_{\rm G}$ and $E(BP-RP)$. The location of the cluster members, both the giant branch and the BSSs, on the CMD is well reproduced by the isochrone.  {The BSSs identified in this work are marked in blue while those re-identified as BSS from the catalogue of Al07 are marked in red. The stars along the giant branch are marked in orange but the 12 giants observed by \citet{Scott95} are marked in green. The six giants with spectrophotometric GDR2 data have been encircled in black. From the isochrone, we find that the stars at the main sequence turn-off have masses $\sim0.9~M_{\odot}$ while the stars along the giant branch have masses $\sim0.9$--$1~M_{\odot}$.
While the evolutionary status of BSSs is uncertain, their single-star masses are inferred as the main sequence turn-off mass corresponding to the PARSEC stellar evolution isochrones with the distance and metallicity appropriate for Be17 but of lower ages. The inferred masses are in the range $\sim1$--$1.6~M_{\odot}$, more than those of the giant branch stars, corresponding to the main sequence turn-off mass of the isochrone with turn-off passing through the faintest and brightest identified BSS respectively.}

\begin{figure}[t]
	\centering
	\includegraphics[width=\columnwidth,angle=0]{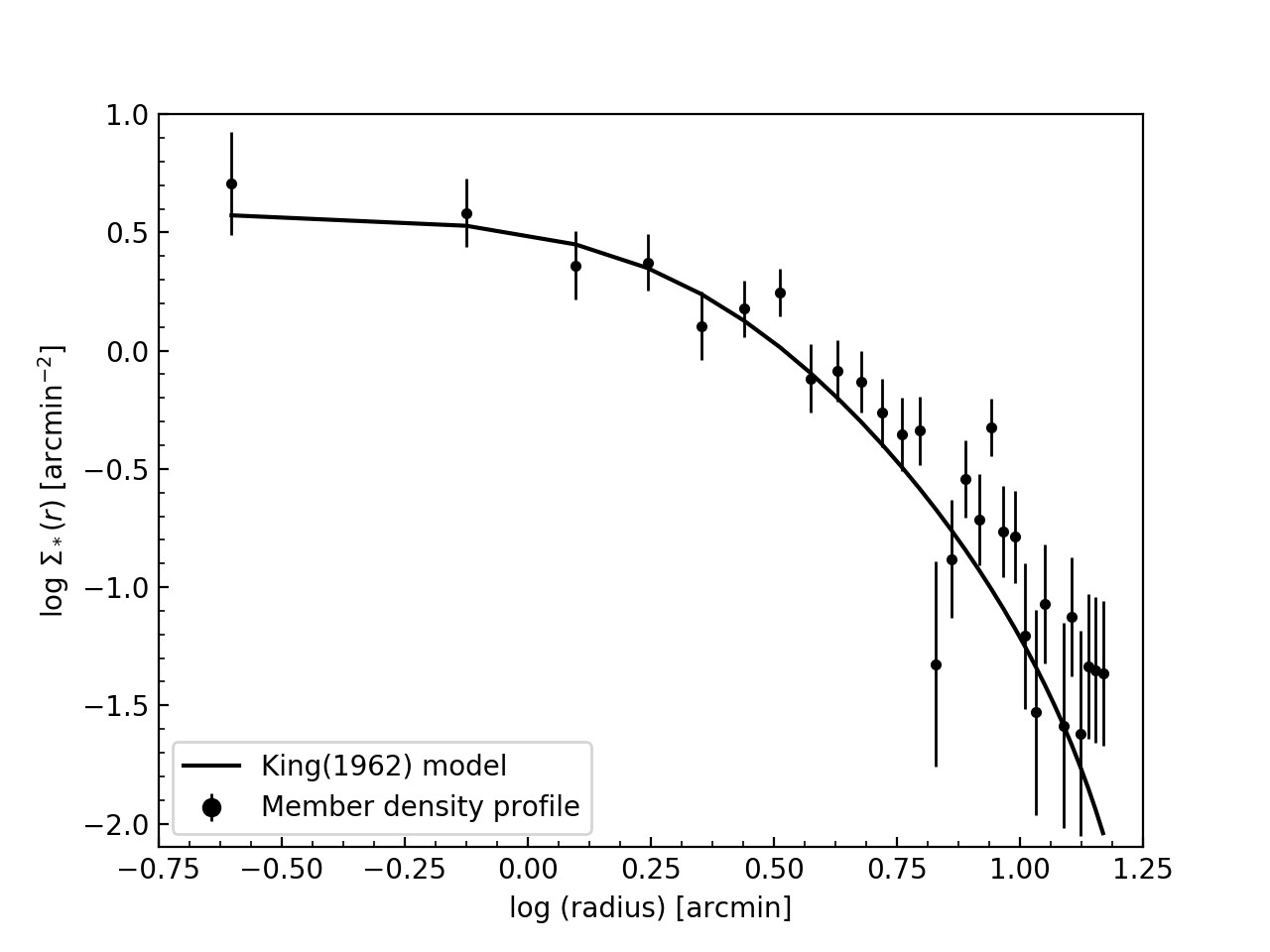}
	\caption{ {Radial density profile of member stars in equi-radial bins fitted with an isotropic single-mass King model given by \citet{king62}.}}
	\label{Fig:king}
\end{figure}

 {Since the radial distribution of BSS (discussed in Sect~\ref{bss}) is highly dependent on the position of the cluster center, we estimate the center of Be17 by finding the centroid of all members in different magnitude bins of 0.5 mag from 15.5~mag to 18~mag, such that each bin has a statistically significant number of cluster members. We obtain the center of Be17 from the mean of the centroid positions of each bin as RA=05:20:33.67, DEC=+30:35:08.71 (J2000) accurate within 15$\arcsec$. We also construct the radial density profile of the cluster by dividing all the observed cluster members into 30 bins at equal radial intervals and computing the number density of member stars in each bin. The resulting profile (as shown in Fig~\ref{Fig:king}) is nicely fitted with an isotropic single-mass King model given by \citet{king62}. The fitting provides the normalization factor (k), core (r$\rm_c$) and tidal (r$\rm_t$) radii as  $4.77 \pm 1.21$ sq. arcmin, $2\arcmin.28 \pm 0\arcmin.75$ and $20\arcmin.76 \pm 9\arcmin.49$ respectively. While the uncertainty is large, within errors the parameters determined are close to those found by \citet{Khar13}, k$= 6.28$ sq. arcmin, r$\rm_c = 1\arcmin.2$ and r$\rm_t = 6\arcmin.94$, estimated without determining membership for Be17. They report errors as negligible but most stars in their observed field were non-members.}

\section{Blue straggler population}
\label{bss}
Interestingly, of the 31 sources identified as BSSs in Be17 by \citet[tabulated in \citealt{Chen17}]{al07}, only two are identified by us as members. The majority of them were already classified as non-members in the proper motion selection (Fig~\ref{Fig:pmsel}) and only two survived the parallax selection (Fig~\ref{Fig:cmd}). This is not surprising because \citet{al07} had not employed any membership criteria in selecting the BSSs and since the background stars of Be17 occupy the same area in the CMD as the BSSs do  \citep{Bh17}, reliable identification of BSSs is not possible without any membership constraints. This effect of field-star contamination on the BSSs identified by \citet{al07} had also been seen by \citet{Car08} for the open clusters NGC\,7789, Berkeley\,66 and Berkeley\,70.

\subsection{Radial distribution}
Being more massive than most cluster members, the BSSs should appear more centrally concentrated in a dynamically evolved star cluster, as a result of mass segregation. Mass segregation is already observed in the core of Be17 \citep{Bh17} showing that it is certainly undergoing dynamical relaxation. The accurate astrometric capabilities of GDR2 enable us identify reliably the stellar population belonging to Be17 from the tip of the red giant branch (RGB) down to the main-sequence turn-off.  {We thus select the BSSs (shown in squares in Fig~\ref{Fig:cmd}) of the cluster, and compare their radial distribution with those along the RGB (shown in orange and green in Fig~\ref{Fig:cmd}), chosen as the reference population, to understand the dynamical evolution of Be17. We identify 23 BSSs in Be17.  In addition we select 45 RGB stars as the reference population with the same magnitude depth as the BSSs, so as to ensure both samples are equally affected by incompleteness.} 

\begin{figure}[t]
	\centering
	\includegraphics[width=\columnwidth,angle=0]{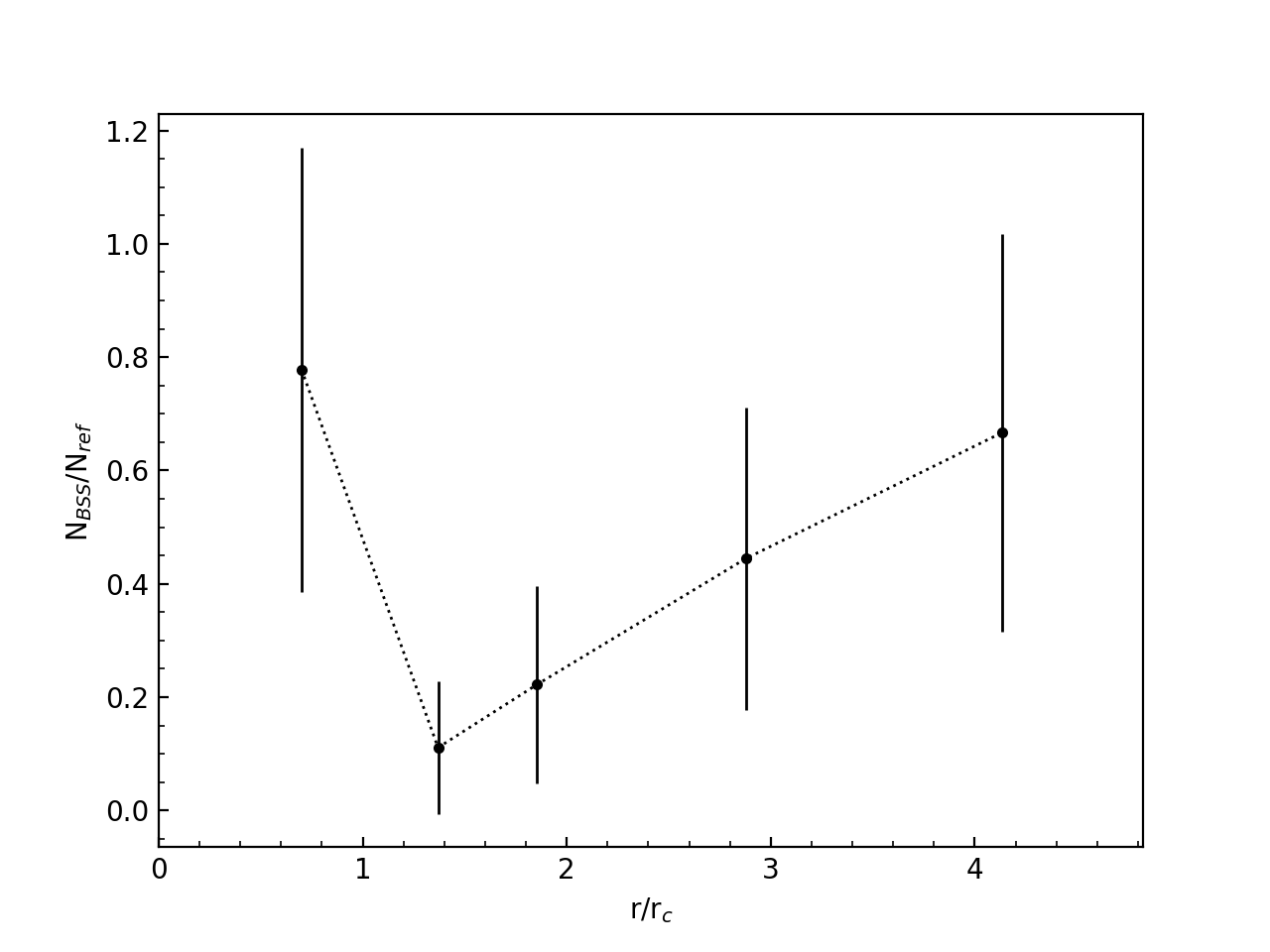}
	\caption{Number of BSSs with respect to that of the reference stars, plotted as a function of the distance from the cluster center expressed in terms of r$\rm_c$.  {The errors are Poisson errors.}}
	\label{Fig:bss_num}
\end{figure}

We first compare the cumulative radial distribution of the BSS population to that of the reference stars to conduct a Kolmogorov–Smirnov test which yields a probability of 99.9\% that the two samples are not extracted from the same parent population. We then divide the area around the cluster into concentric  {circular} annuli such that each annuli has roughly an equal number  {(9 RGBs in each bin)} of reference stars \citep[similar to][]{Lan07,Bec13}.  Fig~\ref{Fig:bss_num} plots the number of BSSs with respect to that of the reference stars in each annulus, as a function of the distance from the cluster center expressed in terms of r$\rm_c$ calculated in Sect~\ref{prop}. It is clearly bimodal, with a peak of BSSs in the inner region, a clear dip at intermediate radii, and an increasing value that is slightly smaller than the central one in the outskirts.  {The minimum occurs at r$_{\rm min} \sim 3.12\arcmin$ or r$_{\rm min}$/r$\rm_c \sim 1.37$. The BSSs are more concentrated in the centre with the central radial bin having 7 BSSs while the second bin with the minima having just one BSS.}

\subsection{Minimum spanning tree}
As an alternative test to evaluate the degree of BSS segregation, in view of the relatively small number of BSSs, we applied the method of the Minimum Spanning Tree \citep[MST;][and references therein]{Alli09}. The MST is the unique set of edges (straight lines) connecting a given sample of vertices (here, the star coordinates) without closed loops, to minimize the sum of the edge lengths.  {The star coordinates are treated as Cartesian points on a plane so the edge lengths are not the same as the distances between the stars on the sky. The length of the MST, $\ell_{\rm MST}$, is the sum of all such edge lengths connecting the vertices. It a measure of the compactness of a given sample of vertices and is independent of the center of the sample. A sample of vertices that is more concentrated on a plane would have a lower $\ell_{\rm MST}$ than a sample of vertices that is more spread out on a plane.} 
We compare the segregation of the BSSs with respect to a reference population, the RGB stars, by their $\ell_{\rm MST}$ lengths \citep[eg.][]{Bec12} within $4\arcmin$ of the cluster center.  {From Fig~\ref{Fig:bss_num}, one would expect that the BSSs are more concentrated than the reference population within this radius and would thus have a lower $\ell_{\rm MST}$ than the reference population. We compute $\Gamma_{\rm MST}$ as the ratio of the MST lengths of the reference population to that of the BSS. We report this dimensionless ratio which would be greater than 1 if the BSS are more concentrated than the reference population. $\ell_{\rm MST}$ is reported in arbitrary units of length.} 

\begin{figure}[t]
	\centering
	\includegraphics[width=\columnwidth,angle=0]{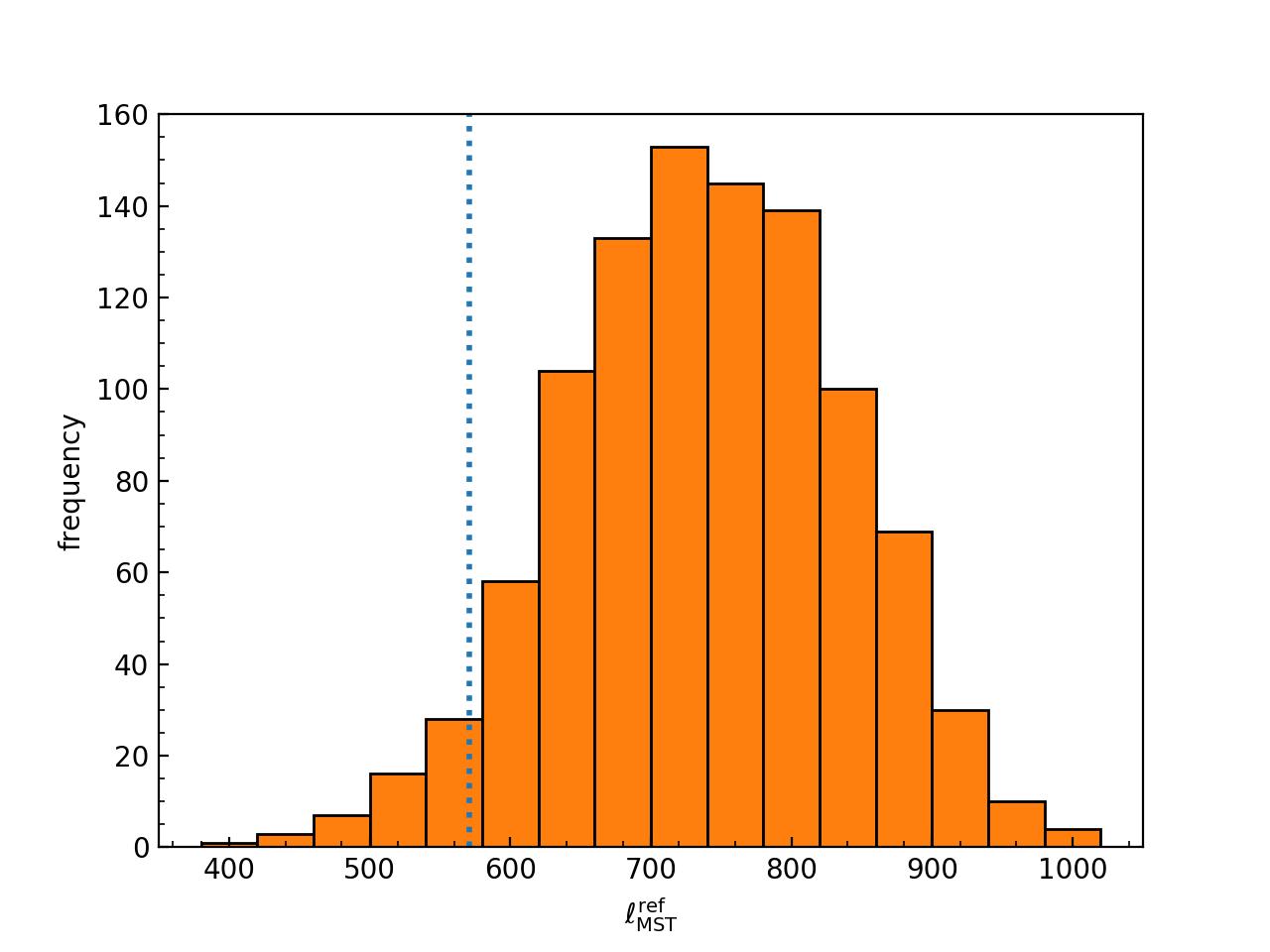}
    \caption{The histogram shows the distribution of the $\ell_{\rm MST}^{\rm ref}$ for the 1000 randomly extracted sets, each of 8 stars, from the reference population. The dotted blue line shows the $\ell_{\rm MST}^{\rm BSS}$.}
	\label{Fig:mst}
\end{figure}

We obtain $\ell_{\rm MST}^{\rm BSS} = 570.8$ for the 8 BSSs located within $4\arcmin$. We then randomly extract 1000 sets of 8 stars from the reference population, and compute the $\ell_{\rm MST}^{\rm ref}$ of each set.  {Its distribution is shown in Fig~\ref{Fig:mst}, from which the mean MST length, $<\ell_{\rm MST}^{\rm ref}> = 736.33$, and standard deviation, $\Delta\ell_{\rm MST}^{\rm ref} = 97.04$ of the distribution, are derived.}  The level of BSS segregation with respect to the reference stars and its associated uncertainty have been estimated as:
$$\Gamma_{\rm MST}= <\ell_{\rm MST}^{\rm ref}>/\ell_{\rm MST}^{\rm BSS}  = 1.29, \Delta\ell_{\rm MST}= \Delta\ell_{\rm MST}^{\rm ref}/\ell_{\rm MST}^{\rm BSS} = 0.17$$
Obtaining a $\Gamma_{\rm MST} > 1$ clearly shows that the BSSs are more concentrated than the reference population within $4\arcmin$. This bolsters the notion that the radial distribution of the BSSs in Be17 can indeed be used as a ``dynamical clock'' for the cluster.

\section{Discussion}
\label{diss}
From the exquisite data available from GDR2, we are able to identify the members of Be17, including the BSSs. Our stringent selection criteria means that we may miss identifying main-sequence members and not the BSS or giant-branch members, both of which should be mostly complete and equally affected by incompleteness. For the brightest cluster members, G < 15 mag, we obtain the mean proper motion of the cluster as $\mu_{\rm \alpha} \cos\delta = 2.59 \pm 0.26$~mas~yr$^{-1}$, $\mu_{\rm \delta} =-0.26 \pm 0.38$~mas~yr$^{-1}$, similar to what we get for the 12 previously-identified giants but offset in $\mu_{\rm \delta}$.  {We observe a spread in the giant branch colour, present also for those stars with GDR2 spectra and hence most certainly members. Such a feature can be caused by a metallicity spread in the giant-branch stars, differential reddening in the cluster, or presence of multiple stellar populations. A significant spread in the metallicity was not observed by \citet{Scott95} for their 12 giants and the cluster, located below the galactic plane towards the galactic anti-center is not expected to show substantial differential reddening enough to cause the spread in colour. Signs of multiple stellar populations have been observed in the open cluster NGC 6791 \citep{Gei12} from Na abundance measurements. Being an old open cluster, we can not rule out the presence of multiple stellar populations in Be17 which may be responsible for the colour spread in the giant-branch stars. Spectroscopic studies of the giant branch stars of Be17 would be required to check for the presence of multiple stellar populations.}

The central concentration of BSSs in the core of Be17 corroborates the massive nature of BSSs compared to the giant branch stars, and hence their suitability as efficient test-particles to infer the dynamical state of stellar clusters. It lends support to the identified BSSs being rejuvenated massive main sequence stars making Be17 a new laboratory to test BSS formation theories with spectroscopic studies. The BSS radial distribution has been used as a powerful tool to shed light on the internal dynamical evolution of globular clusters \citep{Ferraro12,Bec13}. In this work we present the first example of its application for an open cluster. Mass segregation in the core of Be17 \citep{Bh17} already indicated that it was dynamically evolved. Using the radial distribution of the BSSs we can, for the first time, make a direct comparison of the dynamical state of an open cluster with that of globular clusters. The bimodal distribution clearly places Be17 with \textit{Family II} globular clusters, where the BSSs have just started to sink towards the center.  {This is an empirical proof that Be17 is still undergoing dynamical evolution, just like \textit{Family II} globular clusters, which implies that the distribution of stars in Be17 is not dominated by its primordial mass distribution. With a lower stellar density than globular clusters, it is easier to obtain radial velocity measurements of the BSS in Be17. This would allow us to find their trajectories thereby utilizing them as kinematic tracers to better understand its dynamical evolution which would also be characteristic of \textit{Family II} globular clusters.} 

 {It is expected that old open clusters are dynamically evolving and their BSS population, wherever present, can be used to determine their state of dynamical evolution. \citet{al07} indeed observe a significant number of BSS candidates, albeit without reliable membership determination, in many old open clusters whose radial distribution can provide an insight into the dynamical evolution of old open clusters. Since BSSs can only be accurately determined after reliable membership determination, many old open clusters previously classified as being rich in BSSs would need to be re-evaluated with membership determination (Vaidya et al. in preparation). This is important to understand the environmental effects of BSS formation in old open clusters by comparing old open clusters rich in BSS to those that are deficient in BSS \citep[e.g.][]{lee17}.}
 
\begin{acknowledgements}
      The authors are grateful to the anonymous referee for the valuable comments. SB acknowledges support from the IMPRS on Astrophysics at the LMU Munich. This work has made use of data from the European Space Agency (ESA) mission \textit{Gaia}\footnote{https://www.cosmos.esa.int/gaia}, processed by the \textit{Gaia} Data Processing and Analysis Consortium (DPAC\footnote{https://www.cosmos.esa.int/web/gaia/dpac/consortium}). This research made use of Astropy-- a community-developed core Python package for Astronomy \citep{Rob13}, Numpy \citep{numpy} and Matplotlib \citep{matplotlib}. This research also made use of NASA’s Astrophysics Data System (ADS\footnote{https://ui.adsabs.harvard.edu}).
\end{acknowledgements}



\begin{thebibliography}{45}
\expandafter\ifx\csname natexlab\endcsname\relax\def\natexlab#1{#1}\fi

\bibitem[{{Ahumada} \& {Lapasset}(2007)}]{al07}
{Ahumada}, J.~A. \& {Lapasset}, E. 2007, \aap, 463, 789

\bibitem[{{Allison} {et~al.}(2009){Allison}, {Goodwin}, {Parker}, {Portegies
  Zwart}, {de Grijs}, \& {Kouwenhoven}}]{Alli09}
{Allison}, R.~J., {Goodwin}, S.~P., {Parker}, R.~J., {et~al.} 2009, \mnras,
  395, 1449

\bibitem[{{Astropy Collaboration} {et~al.}(2013){Astropy Collaboration},
  {Robitaille}, {Tollerud}, {Greenfield}, {Droettboom}, {Bray}, {Aldcroft},
  {Davis}, {Ginsburg}, {Price-Whelan}, {Kerzendorf}, {Conley}, {Crighton},
  {Barbary}, {Muna}, {Ferguson}, {Grollier}, {Parikh}, {Nair}, {Unther},
  {Deil}, {Woillez}, {Conseil}, {Kramer}, {Turner}, {Singer}, {Fox}, {Weaver},
  {Zabalza}, {Edwards}, {Azalee Bostroem}, {Burke}, {Casey}, {Crawford},
  {Dencheva}, {Ely}, {Jenness}, {Labrie}, {Lim}, {Pierfederici}, {Pontzen},
  {Ptak}, {Refsdal}, {Servillat}, \& {Streicher}}]{Rob13}
{Astropy Collaboration}, {Robitaille}, T.~P., {Tollerud}, E.~J., {et~al.} 2013,
  \aap, 558, A33

\bibitem[{{Bailer-Jones} {et~al.}(2013){Bailer-Jones}, {Andrae}, {Arcay},
  {Astraatmadja}, {Bellas-Velidis}, {Berihuete}, {Bijaoui}, {Carri{\'o}n},
  {Dafonte}, {Damerdji}, {Dapergolas}, {de Laverny}, {Delchambre}, {Drazinos},
  {Drimmel}, {Fr{\'e}mat}, {Fustes}, {Garc{\'\i}a-Torres}, {Gu{\'e}d{\'e}},
  {Heiter}, {Janotto}, {Karampelas}, {Kim}, {Knude}, {Kolka}, {Kontizas},
  {Kontizas}, {Korn}, {Lanzafame}, {Lebreton}, {Lindstr{\o}m}, {Liu},
  {Livanou}, {Lobel}, {Manteiga}, {Martayan}, {Ordenovic}, {Pichon},
  {Recio-Blanco}, {Rocca-Volmerange}, {Sarro}, {Smith}, {Sordo}, {Soubiran},
  {Surdej}, {Th{\'e}venin}, {Tsalmantza}, {Vallenari}, \& {Zorec}}]{bj13}
{Bailer-Jones}, C.~A.~L., {Andrae}, R., {Arcay}, B., {et~al.} 2013, \aap, 559,
  A74

\bibitem[{{Bailer-Jones} {et~al.}(2018){Bailer-Jones}, {Rybizki}, {Fouesneau},
  {Mantelet}, \& {Andrae}}]{bj18}
{Bailer-Jones}, C.~A.~L., {Rybizki}, J., {Fouesneau}, M., {Mantelet}, G., \&
  {Andrae}, R. 2018, \aj, 156, 58

\bibitem[{{Beccari} {et~al.}(2013){Beccari}, {Dalessandro}, {Lanzoni},
  {Ferraro}, {Sollima}, {Bellazzini}, \& {Miocchi}}]{Bec13}
{Beccari}, G., {Dalessandro}, E., {Lanzoni}, B., {et~al.} 2013, \apj, 776, 60

\bibitem[{{Beccari} {et~al.}(2012){Beccari}, {L{\"u}tzgendorf}, {Olczak},
  {Ferraro}, {Lanzoni}, {Carraro}, {Stetson}, {Sollima}, \& {Boffin}}]{Bec12}
{Beccari}, G., {L{\"u}tzgendorf}, N., {Olczak}, C., {et~al.} 2012, \apj, 754,
  108

\bibitem[{{Bertelli Motta} {et~al.}(2018){Bertelli Motta}, {Pasquali},
  {Caffau}, \& {Grebel}}]{Bertelli18}
{Bertelli Motta}, C., {Pasquali}, A., {Caffau}, E., \& {Grebel}, E.~K. 2018,
  \mnras, 480, 4314

\bibitem[{{Bhattacharya} {et~al.}(2017){Bhattacharya}, {Mishra}, {Vaidya}, \&
  {Chen}}]{Bh17}
{Bhattacharya}, S., {Mishra}, I., {Vaidya}, K., \& {Chen}, W.~P. 2017, \apj,
  847, 138

\bibitem[{{Bragaglia} {et~al.}(2006){Bragaglia}, {Tosi}, {Andreuzzi}, \&
  {Marconi}}]{Bragaglia06}
{Bragaglia}, A., {Tosi}, M., {Andreuzzi}, G., \& {Marconi}, G. 2006, \mnras,
  368, 1971

\bibitem[{{Bressan} {et~al.}(2012){Bressan}, {Marigo}, {Girardi}, {Salasnich},
  {Dal Cero}, {Rubele}, \& {Nanni}}]{bre12}
{Bressan}, A., {Marigo}, P., {Girardi}, L., {et~al.} 2012, \mnras, 427, 127

\bibitem[{{Cantat-Gaudin} {et~al.}(2018){Cantat-Gaudin}, {Jordi}, {Vallenari},
  {Bragaglia}, {Balaguer-N{\'u}{\~n}ez}, {Soubiran}, {Bossini}, {Moitinho},
  {Castro-Ginard}, {Krone-Martins}, {Casamiquela}, {Sordo}, \&
  {Carrera}}]{cg18}
{Cantat-Gaudin}, T., {Jordi}, C., {Vallenari}, A., {et~al.} 2018, ArXiv
  e-prints, arXiv:1805.08726

\bibitem[{{Carraro} {et~al.}(2008){Carraro}, {V{\'a}zquez}, \&
  {Moitinho}}]{Car08}
{Carraro}, G., {V{\'a}zquez}, R.~A., \& {Moitinho}, A. 2008, \aap, 482, 777

\bibitem[{{Chen} {et~al.}(2017){Chen}, {Bhattacharya}, {Mishra}, {Vaidya}, \&
  {Lalchand}}]{Chen17}
{Chen}, W.~P., {Bhattacharya}, S., {Mishra}, I., {Vaidya}, K., \& {Lalchand},
  B. 2017, in Journal of Physics Conference Series, Vol. 869, 012093

\bibitem[{{Chen} {et~al.}(2004){Chen}, {Chen}, \& {Shu}}]{Chen04}
{Chen}, W.~P., {Chen}, C.~W., \& {Shu}, C.~G. 2004, \aj, 128, 2306

\bibitem[{{Chen} {et~al.}(2014){Chen}, {Girardi}, {Bressan}, {Marigo},
  {Barbieri}, \& {Kong}}]{che14}
{Chen}, Y., {Girardi}, L., {Bressan}, A., {et~al.} 2014, \mnras, 444, 2525

\bibitem[{{Dalessandro} {et~al.}(2009){Dalessandro}, {Beccari}, {Lanzoni},
  {Ferraro}, {Schiavon}, \& {Rood}}]{Dal09}
{Dalessandro}, E., {Beccari}, G., {Lanzoni}, B., {et~al.} 2009, \apjs, 182, 509

\bibitem[{{Davies} {et~al.}(2004){Davies}, {Piotto}, \& {de Angeli}}]{Davies04}
{Davies}, M.~B., {Piotto}, G., \& {de Angeli}, F. 2004, \mnras, 349, 129

\bibitem[{{Dias} {et~al.}(2014){Dias}, {Monteiro}, {Caetano}, {L{\'e}pine},
  {Assafin}, \& {Oliveira}}]{dia14}
{Dias}, W.~S., {Monteiro}, H., {Caetano}, T.~C., {et~al.} 2014, \aap, 564, A79

\bibitem[{{Evans} {et~al.}(2018){Evans}, {Riello}, {De Angeli}, {Carrasco},
  {Montegriffo}, {Fabricius}, {Jordi}, {Palaversa}, {Diener}, {Busso},
  {Cacciari}, {van Leeuwen}, {Burgess}, {Davidson}, {Harrison}, {Hodgkin},
  {Pancino}, {Richards}, {Altavilla}, {Balaguer-N{\'u}{\~n}ez}, {Barstow},
  {Bellazzini}, {Brown}, {Castellani}, {Cocozza}, {De Luise}, {Delgado},
  {Ducourant}, {Galleti}, {Gilmore}, {Giuffrida}, {Holl}, {Kewley}, {Koposov},
  {Marinoni}, {Marrese}, {Osborne}, {Piersimoni}, {Portell}, {Pulone},
  {Ragaini}, {Sanna}, {Terrett}, {Walton}, {Wevers}, \&
  {Wyrzykowski}}]{Evans18}
{Evans}, D.~W., {Riello}, M., {De Angeli}, F., {et~al.} 2018, \aap, 616, A4

\bibitem[{{Ferraro} {et~al.}(2009){Ferraro}, {Beccari}, {Dalessandro},
  {Lanzoni}, {Sills}, {Rood}, {Pecci}, {Karakas}, {Miocchi}, \&
  {Bovinelli}}]{Ferraro09}
{Ferraro}, F.~R., {Beccari}, G., {Dalessandro}, E., {et~al.} 2009, \nat, 462,
  1028

\bibitem[{{Ferraro} {et~al.}(2012){Ferraro}, {Lanzoni}, {Dalessandro},
  {Beccari}, {Pasquato}, {Miocchi}, {Rood}, {Sigurdsson}, {Sills}, {Vesperini},
  {Mapelli}, {Contreras}, {Sanna}, \& {Mucciarelli}}]{Ferraro12}
{Ferraro}, F.~R., {Lanzoni}, B., {Dalessandro}, E., {et~al.} 2012, \nat, 492,
  393

\bibitem[{{Friel} {et~al.}(2002){Friel}, {Janes}, {Tavarez}, {Scott},
  {Katsanis}, {Lotz}, {Hong}, \& {Miller}}]{Friel02}
{Friel}, E.~D., {Janes}, K.~A., {Tavarez}, M., {et~al.} 2002, \aj, 124, 2693

\bibitem[{{Gaia Collaboration} {et~al.}(2018){Gaia Collaboration}, {Brown},
  {Vallenari}, {Prusti}, {de Bruijne}, {Babusiaux}, {Bailer-Jones}, {Biermann},
  {Evans}, {Eyer}, {Jansen}, {Jordi}, {Klioner}, {Lammers}, {Lindegren},
  {Luri}, {Mignard}, {Panem}, {Pourbaix}, {Randich}, {Sartoretti}, {Siddiqui},
  {Soubiran}, {van Leeuwen}, {Walton}, {Arenou}, {Bastian}, {Cropper},
  {Drimmel}, {Katz}, {Lattanzi}, {Bakker}, {Cacciari}, {Casta{\~n}eda},
  {Chaoul}, {Cheek}, {De Angeli}, {Fabricius}, {Guerra}, {Holl}, {Masana},
  {Messineo}, {Mowlavi}, {Nienartowicz}, {Panuzzo}, {Portell}, {Riello},
  {Seabroke}, {Tanga}, {Th{\'e}venin}, {Gracia-Abril}, {Comoretto},
  {Garcia-Reinaldos}, {Teyssier}, {Altmann}, {Andrae}, {Audard},
  {Bellas-Velidis}, {Benson}, {Berthier}, {Blomme}, {Burgess}, {Busso},
  {Carry}, {Cellino}, {Clementini}, {Clotet}, {Creevey}, {Davidson}, {De
  Ridder}, {Delchambre}, {Dell'Oro}, {Ducourant},
  {Fern{\'a}ndez-Hern{\'a}ndez}, {Fouesneau}, {Fr{\'e}mat}, {Galluccio},
  {Garc{\'\i}a-Torres}, {Gonz{\'a}lez-N{\'u}{\~n}ez}, {Gonz{\'a}lez- Vidal},
  {Gosset}, {Guy}, {Halbwachs}, {Hambly}, {Harrison}, {Hern{\'a}ndez},
  {Hestroffer}, {Hodgkin}, {Hutton}, {Jasniewicz}, {Jean-Antoine- Piccolo},
  {Jordan}, {Korn}, {Krone- Martins}, {Lanzafame}, {Lebzelter}, {L{\"o}ffler},
  {Manteiga}, {Marrese}, {Mart{\'\i}n-Fleitas}, {Moitinho}, {Mora}, {Muinonen},
  {Osinde}, {Pancino}, {Pauwels}, {Petit}, {Recio-Blanco}, {Richards},
  {Rimoldini}, {Robin}, {Sarro}, {Siopis}, {Smith}, {Sozzetti}, {S{\"u}veges},
  {Torra}, {van Reeven}, {Abbas}, {Abreu Aramburu}, {Accart}, {Aerts},
  {Altavilla}, {{\'A}lvarez}, {Alvarez}, {Alves}, {Anderson}, {Andrei},
  {Anglada Varela}, {Antiche}, {Antoja}, {Arcay}, {Astraatmadja}, {Bach},
  {Baker}, {Balaguer-N{\'u}{\~n}ez}, {Balm}, {Barache}, {Barata}, {Barbato},
  {Barblan}, {Barklem}, {Barrado}, {Barros}, {Barstow}, {Bartholom{\'e}
  Mu{\~n}oz}, {Bassilana}, {Becciani}, {Bellazzini}, {Berihuete}, {Bertone},
  {Bianchi}, {Bienaym{\'e}}, {Blanco-Cuaresma}, {Boch}, {Boeche}, {Bombrun},
  {Borrachero}, {Bossini}, {Bouquillon}, {Bourda}, {Bragaglia}, {Bramante},
  {Breddels}, {Bressan}, {Brouillet}, {Br{\"u}semeister}, {Brugaletta},
  {Bucciarelli}, {Burlacu}, {Busonero}, {Butkevich}, {Buzzi}, {Caffau},
  {Cancelliere}, {Cannizzaro}, {Cantat-Gaudin}, {Carballo}, {Carlucci},
  {Carrasco}, {Casamiquela}, {Castellani}, {Castro-Ginard}, {Charlot},
  {Chemin}, {Chiavassa}, {Cocozza}, {Costigan}, {Cowell}, {Crifo}, {Crosta},
  {Crowley}, {Cuypers}, {Dafonte}, {Damerdji}, {Dapergolas}, {David}, {David},
  {de Laverny}, {De Luise}, {De March}, {de Martino}, {de Souza}, {de Torres},
  {Debosscher}, {del Pozo}, {Delbo}, {Delgado}, {Delgado}, {Di Matteo},
  {Diakite}, {Diener}, {Distefano}, {Dolding}, {Drazinos}, {Dur{\'a}n},
  {Edvardsson}, {Enke}, {Eriksson}, {Esquej}, {Eynard Bontemps}, {Fabre},
  {Fabrizio}, {Faigler}, {Falc{\~a}o}, {Farr{\`a}s Casas}, {Federici},
  {Fedorets}, {Fernique}, {Figueras}, {Filippi}, {Findeisen}, {Fonti},
  {Fraile}, {Fraser}, {Fr{\'e}zouls}, {Gai}, {Galleti}, {Garabato},
  {Garc{\'\i}a-Sedano}, {Garofalo}, {Garralda}, {Gavel}, {Gavras}, {Gerssen},
  {Geyer}, {Giacobbe}, {Gilmore}, {Girona}, {Giuffrida}, {Glass}, {Gomes},
  {Granvik}, {Gueguen}, {Guerrier}, {Guiraud}, {Guti{\'e}rrez-S{\'a}nchez},
  {Haigron}, {Hatzidimitriou}, {Hauser}, {Haywood}, {Heiter}, {Helmi}, {Heu},
  {Hilger}, {Hobbs}, {Hofmann}, {Holland}, {Huckle}, {Hypki}, {Icardi},
  {Jan{\ss}en}, {Jevardat de Fombelle}, {Jonker}, {Juh{\'a}sz}, {Julbe},
  {Karampelas}, {Kewley}, {Klar}, {Kochoska}, {Kohley}, {Kolenberg},
  {Kontizas}, {Kontizas}, {Koposov}, {Kordopatis}, {Kostrzewa-Rutkowska},
  {Koubsky}, {Lambert}, {Lanza}, {Lasne}, {Lavigne}, {Le Fustec}, {Le
  Poncin-Lafitte}, {Lebreton}, {Leccia}, {Leclerc}, {Lecoeur-Taibi},
  {Lenhardt}, {Leroux}, {Liao}, {Licata}, {Lindstr{\o}m}, {Lister}, {Livanou},
  {Lobel}, {L{\'o}pez}, {Managau}, {Mann}, {Mantelet}, {Marchal}, {Marchant},
  {Marconi}, {Marinoni}, {Marschalk{\'o}}, {Marshall}, {Martino}, {Marton},
  {Mary}, {Massari}, {Matijevi{\v{c}}}, {Mazeh}, {McMillan}, {Messina},
  {Michalik}, {Millar}, {Molina}, {Molinaro}, {Moln{\'a}r}, {Montegriffo},
  {Mor}, {Morbidelli}, {Morel}, {Morris}, {Mulone}, {Muraveva}, {Musella},
  {Nelemans}, {Nicastro}, {Noval}, {O'Mullane}, {Ord{\'e}novic},
  {Ord{\'o}{\~n}ez-Blanco}, {Osborne}, {Pagani}, {Pagano}, {Pailler},
  {Palacin}, {Palaversa}, {Panahi}, {Pawlak}, {Piersimoni}, {Pineau}, {Plachy},
  {Plum}, {Poggio}, {Poujoulet}, {Pr{\v{s}}a}, {Pulone}, {Racero}, {Ragaini},
  {Rambaux}, {Ramos-Lerate}, {Regibo}, {Reyl{\'e}}, {Riclet}, {Ripepi}, {Riva},
  {Rivard}, {Rixon}, {Roegiers}, {Roelens}, {Romero-G{\'o}mez}, {Rowell},
  {Royer}, {Ruiz-Dern}, {Sadowski}, {Sagrist{\`a} Sell{\'e}s}, {Sahlmann},
  {Salgado}, {Salguero}, {Sanna}, {Santana- Ros}, {Sarasso}, {Savietto},
  {Schultheis}, {Sciacca}, {Segol}, {Segovia}, {S{\'e}gransan}, {Shih},
  {Siltala}, {Silva}, {Smart}, {Smith}, {Solano}, {Solitro}, {Sordo}, {Soria
  Nieto}, {Souchay}, {Spagna}, {Spoto}, {Stampa}, {Steele},
  {Steidelm{\"u}ller}, {Stephenson}, {Stoev}, {Suess}, {Surdej}, {Szabados},
  {Szegedi-Elek}, {Tapiador}, {Taris}, {Tauran}, {Taylor}, {Teixeira},
  {Terrett}, {Teyssandier}, {Thuillot}, {Titarenko}, {Torra Clotet}, {Turon},
  {Ulla}, {Utrilla}, {Uzzi}, {Vaillant}, {Valentini}, {Valette}, {van Elteren},
  {Van Hemelryck}, {van Leeuwen}, {Vaschetto}, {Vecchiato}, {Veljanoski},
  {Viala}, {Vicente}, {Vogt}, {von Essen}, {Voss}, {Votruba}, {Voutsinas},
  {Walmsley}, {Weiler}, {Wertz}, {Wevers}, {Wyrzykowski}, {Yoldas},
  {{\v{Z}}erjal}, {Ziaeepour}, {Zorec}, {Zschocke}, {Zucker}, {Zurbach}, \&
  {Zwitter}}]{Gaia18}
{Gaia Collaboration}, {Brown}, A.~G.~A., {Vallenari}, A., {et~al.} 2018, \aap,
  616, A1

\bibitem[{{Geisler} {et~al.}(2012){Geisler}, {Villanova}, {Carraro},
  {Pilachowski}, {Cummings}, {Johnson}, \& {Bresolin}}]{Gei12}
{Geisler}, D., {Villanova}, S., {Carraro}, G., {et~al.} 2012, \apj, 756, L40

\bibitem[{{Hills} \& {Day}(1976)}]{Hills76}
{Hills}, J.~G. \& {Day}, C.~A. 1976, Astrophysical Letters, 17, 87

\bibitem[{Hunter(2007)}]{matplotlib}
Hunter, J.~D. 2007, Computing In Science \& Engineering, 9, 90

\bibitem[{{Kaluzny}(1994)}]{kal94}
{Kaluzny}, J. 1994, \actaa, 44, 247

\bibitem[{{Katz} {et~al.}(2018){Katz}, {Sartoretti}, {Cropper}, {Panuzzo},
  {Seabroke}, {Viala}, {Benson}, {Blomme}, {Jasniewicz}, {Jean-Antoine},
  {Huckle}, {Smith}, {Baker}, {Crifo}, {Damerdji}, {David}, {Dolding},
  {Fr{\'e}mat}, {Gosset}, {Guerrier}, {Guy}, {Haigron}, {Jan{\ss}en},
  {Marchal}, {Plum}, {Soubiran}, {Th{\'e}venin}, {Ajaj}, {Allende Prieto},
  {Babusiaux}, {Boudreault}, {Chemin}, {Delle Luche}, {Fabre}, {Gueguen},
  {Hambly}, {Lasne}, {Meynadier}, {Pailler}, {Panem}, {Royer}, {Tauran},
  {Zurbach}, {Zwitter}, {Arenou}, {Bossini}, {Gomez}, {Lemaitre}, {Leclerc},
  {Morel}, {Munari}, {Turon}, {Vallenari}, \& {{\v Z}erjal}}]{Katz18}
{Katz}, D., {Sartoretti}, P., {Cropper}, M., {et~al.} 2018, ArXiv e-prints
  [\eprint[arXiv]{1804.09372}]

\bibitem[{{Kharchenko} {et~al.}(2013){Kharchenko}, {Piskunov}, {Schilbach},
  {R{\"o}ser}, \& {Scholz}}]{Khar13}
{Kharchenko}, N.~V., {Piskunov}, A.~E., {Schilbach}, E., {R{\"o}ser}, S., \&
  {Scholz}, R.~D. 2013, \aap, 558, A53

\bibitem[{{King}(1962)}]{king62}
{King}, I. 1962, \aj, 67, 471

\bibitem[{{Krusberg} \& {Chaboyer}(2006)}]{kru06}
{Krusberg}, Z. A.~C. \& {Chaboyer}, B. 2006, \aj, 131, 1565

\bibitem[{{Lanzoni} {et~al.}(2007){Lanzoni}, {Dalessandro}, {Ferraro},
  {Mancini}, {Beccari}, {Rood}, {Mapelli}, \& {Sigurdsson}}]{Lan07}
{Lanzoni}, B., {Dalessandro}, E., {Ferraro}, F.~R., {et~al.} 2007, \apj, 663,
  267

\bibitem[{{Lee} \& {Chang}(2017)}]{lee17}
{Lee}, H.-U. \& {Chang}, H.-Y. 2017, Journal of Korean Astronomical Society,
  50, 51

\bibitem[{{Lindegren} {et~al.}(2018){Lindegren}, {Hern{\'a}ndez}, {Bombrun},
  {Klioner}, {Bastian}, {Ramos-Lerate}, {de Torres}, {Steidelm{\"u}ller},
  {Stephenson}, {Hobbs}, {Lammers}, {Biermann}, {Geyer}, {Hilger}, {Michalik},
  {Stampa}, {McMillan}, {Casta{\~n}eda}, {Clotet}, {Comoretto}, {Davidson},
  {Fabricius}, {Gracia}, {Hambly}, {Hutton}, {Mora}, {Portell}, {van Leeuwen},
  {Abbas}, {Abreu}, {Altmann}, {Andrei}, {Anglada}, {Balaguer-N{\'u}{\~n}ez},
  {Barache}, {Becciani}, {Bertone}, {Bianchi}, {Bouquillon}, {Bourda},
  {Br{\"u}semeister}, {Bucciarelli}, {Busonero}, {Buzzi}, {Cancelliere},
  {Carlucci}, {Charlot}, {Cheek}, {Crosta}, {Crowley}, {de Bruijne}, {de
  Felice}, {Drimmel}, {Esquej}, {Fienga}, {Fraile}, {Gai}, {Garralda},
  {Gonz{\'a}lez- Vidal}, {Guerra}, {Hauser}, {Hofmann}, {Holl}, {Jordan},
  {Lattanzi}, {Lenhardt}, {Liao}, {Licata}, {Lister}, {L{\"o}ffler},
  {Marchant}, {Martin-Fleitas}, {Messineo}, {Mignard}, {Morbidelli}, {Poggio},
  {Riva}, {Rowell}, {Salguero}, {Sarasso}, {Sciacca}, {Siddiqui}, {Smart},
  {Spagna}, {Steele}, {Taris}, {Torra}, {van Elteren}, {van Reeven}, \&
  {Vecchiato}}]{Lin18}
{Lindegren}, L., {Hern{\'a}ndez}, J., {Bombrun}, A., {et~al.} 2018, \aap, 616,
  A2

\bibitem[{{Marigo} {et~al.}(2017){Marigo}, {Girardi}, {Bressan}, {Rosenfield},
  {Aringer}, {Chen}, {Dussin}, {Nanni}, {Pastorelli}, {Rodrigues}, {Trabucchi},
  {Bladh}, {Dalcanton}, {Groenewegen}, {Montalb{\'a}n}, \& {Wood}}]{marigo17}
{Marigo}, P., {Girardi}, L., {Bressan}, A., {et~al.} 2017, \apj, 835, 77

\bibitem[{{Mathieu} \& {Geller}(2009)}]{mg09}
{Mathieu}, R.~D. \& {Geller}, A.~M. 2009, \nat, 462, 1032

\bibitem[{{McCrea}(1964)}]{McCrea64}
{McCrea}, W.~H. 1964, \mnras, 128, 147

\bibitem[{Oliphant(2015)}]{numpy}
Oliphant, T.~E. 2015, Guide to NumPy, 2nd edn. (USA: CreateSpace Independent
  Publishing Platform)

\bibitem[{{Phelps}(1997)}]{Phelps97}
{Phelps}, R.~L. 1997, \apj, 483, 826

\bibitem[{{Salaris} {et~al.}(2004){Salaris}, {Weiss}, \& {Percival}}]{sal04}
{Salaris}, M., {Weiss}, A., \& {Percival}, S.~M. 2004, \aap, 414, 163

\bibitem[{{Sandage}(1953)}]{sandage53}
{Sandage}, A.~R. 1953, \aj, 58, 61

\bibitem[{{Scott} {et~al.}(1995){Scott}, {Friel}, \& {Janes}}]{Scott95}
{Scott}, J.~E., {Friel}, E.~D., \& {Janes}, K.~A. 1995, \aj, 109, 1706

\bibitem[{{Sills} {et~al.}(2013){Sills}, {Glebbeek}, {Chatterjee}, \&
  {Rasio}}]{Sills13}
{Sills}, A., {Glebbeek}, E., {Chatterjee}, S., \& {Rasio}, F.~A. 2013, \apj,
  777, 105

\bibitem[{{Yen} {et~al.}(2018){Yen}, {Reffert}, {Schilbach}, {R{\"o}ser},
  {Kharchenko}, \& {Piskunov}}]{Yen18}
{Yen}, S.~X., {Reffert}, S., {Schilbach}, E., {et~al.} 2018, \aap, 615, A12

\end{thebibliography}
\end{document}